\newcommand{\bvec}[1]{{\bf #1}}
\documentstyle[prl,aps,graphicx,multicol]{revtex}
\begin{document}
\draft

\title{Solving the crystallographic phase problem in i(AlPdMn)}
\author{K. S. Brown, A. A. Avanesov\cite{sasha}, and V. Elser}
\address{Laboratory of Atomic and Solid State Physics, Clark Hall, Cornell
University, Ithaca, NY 14853-2501}
\date{\today}

\maketitle
\begin{abstract}

We apply a new technique for \emph{ab initio} phase determination
[Acta Crystallogr.~Sect.~A, A{\bf 55}, 48
(1999)] to solve for the average structure of the icosahedral (i) phase
of AlPdMn. After an introduction to the crystallographic phase
problem and a description of the method, we present a brief
report of our findings for the structure of i(AlPdMn).  Despite
the use of data from extremely high quality samples, we find
strong evidence of disorder in the
structure, lending support to the random tiling model of quasicrystal
stabilization.

\end{abstract}


\begin{multicols}{2}
While X--ray diffraction is one of the most useful tools we have
for solid--state structure determination for everything from
simple metals to protein crystals, in a very real sense
diffraction experiments only provide half of the information we
need.  X--ray diffractometers measure only the amplitudes of the
various Fourier components of the electron density, not their
relative phases. The recovery of the phases, given only the
amplitudes, is termed the ``phase problem'', or ``phase
retrieval'', in a more general context.  Efforts to solve the
phase problem can be broadly grouped into those that use
additional data and those that do not. In the first category are
three--beam interference measurements\cite{ref:threeBeam} and
techniques such as isomorphous substitution and anomalous
scattering at multiple wavelengths
\cite{ref:drenth,ref:anomalous}.   Methods of the second type are
termed ``direct'' methods and were pioneered by Hauptman and
Karle \cite{ref:H&K}.  Algorithms which implement direct methods
are highly evolved and very successful in solving structures of
up to a few hundred atoms \cite{ref:sheldrick}. Recently, a
technique called ``Shake 'n Bake'' (SnB)
\cite{ref:SnBweb,ref:SnBpaper} was developed expressly for
application to more complex structures such as proteins. Common
to all direct method algorithms currently in use is an iteration
scheme which modifies phases to reproduce, either the statistical
distribution of phase invariants (triplets, etc.) expected of an
ideal gas of equal point scatterers, or known general features of
the electron density (e.g. ``solvent flattening''). Due to their
iterative nature, the success rate of these algorithms, starting
from random phases, is very unpredictable and in the case of
complex structures frequently only a small fraction of a
percent\cite{ref:SnBweb}. One possible reason for this behavior
is that like all iterative optimization schemes
\cite{ref:annealcomment} these methods can become stuck in fixed
points that are locally but not globally optimal.  This
phenomenon is known as ``stagnation'' in the image processing
literature \cite{ref:images}.

Unfortunately, current direct methods are completely unsuitable
for solving i(AlPdMn) and all other quasicrystalline structures
for two reasons.  On the one hand, the properties of the
quasicrystalline state require that their Bragg peaks be indexed
by more than three indices; icosahedral phases, in
particular, have an average structure that requires six dimensions
(6D) to
represent periodically\cite{ref:qc}. Furthermore, an atomistic
three dimensional electron density is obtained by cutting a density in
the higher dimensional space. The latter need not be concentrated at
points; rather, one expects the analogues of atoms --- ``atomic
surfaces'' --- to be rather
extended in the additional dimensions. The inability of
established direct methods to deal with these features of
quasicrystalline charge distributions recently led one of us
\cite{ref:mincharge} to introduce a new method for phase
retrieval, called the principle
of minimum charge (minQ).  This new method differs from previous
direct methods in three important respects: it is not iterative
or stochastic, it does not
rely on atomicity of the charge density, and it is applicable in
arbitrary dimension. A direct method that works in arbitrary
dimension is useful not only for quasicrystals but also in image
reconstruction.

While an extended discussion of the foundations of minQ is
available elsewhere \cite{ref:mincharge},
the central idea can be stated very simply: \emph{the correct
phases are such that the total charge $Q$ is a minimum}. This
claim has its roots in the following observation. The minimum average
charge density (equivalently, $Q$) required to restore
positivity to a tentative density reconstruction is determined by the
deepest minimum in the electron density; the large number of
free parameters (phases) implies
that the minimum $Q$ is achieved when a correspondingly large number
of minima in the electron density are exactly degenerate and represents
the relatively large fraction of the unit cell volume --- in a
correct reconstruction --- having negligible
charge density.

Past methods to solve the phase problem in quasicrystals
have been indirect and have relied on observations about the
diffracted intensities \cite{ref:qperp} or relationships of the
actual structure to closely related approximant structures
\cite{ref:approx}.  There has also been some use of the
maximum--entropy method \cite{ref:maxS} in quasicrystal structure
determination \cite{ref:qcmaxS}, but this work has focused on
structure refinement and not \emph{ab initio} phase recovery. Our selection of
i(AlPdMn) as a proving ground for the minQ method had a threefold
motivation: as previously stated, existing direct methods are not
applicable, important structural details were heretofore unknown,
and a wealth of
diffraction data for extremely high quality samples is available
\cite{ref:boudard}.

The Fourier series for the
electron density, for crystals as well as quasicrystals, may be
written as
\begin{equation}
\rho(\bvec{r}) = F_{\bvec{0}} + 2\sum_{\bvec{q}\in\Lambda^{+}} F_{\bvec{q}}
\cos(\bvec{q} \cdot \bvec{r} - \phi_{\bvec{q}})
\label{eqn:seriesone}
\end{equation}
where $\Lambda^{+}$ is a set of nonzero reciprocal lattice vectors
corresponding to the x-ray data
(each element representing a
$\{\bvec{q},-\bvec{q}\}$ pair). $F_{\bvec{0}}$ is the average charge density
whose minimization is the goal of any algorithm based on minQ.
We assume the size of $\Lambda^{+}$ is large enough so as
to make truncation errors negligible. If we specialize to the
case of structures with centrosymmetry, as in i(AlPdMn),
$\rho(\bvec{r}) = \rho(-\bvec{r})$ and $\phi_{\bvec{q}}$ can be
either 0 or $\pi$.  We may therefore replace the phases in the
argument of the cosine with overall multiplicative signs
$s_{\bvec{q}} =\pm 1$.  Our separation of $F_{\bvec{0}}$ from the rest of
the diffraction amplitudes in Eqn.~(\ref{eqn:seriesone})
emphasizes its practical difference from other amplitudes in that
$F_{\bvec{0}}$ is not measured. Given a maximum spatial frequency in
$\Lambda^{+}$, we can evaluate the density on a finite grid
$\mathcal{G}$.
Consequently, we only deal with a matrix of numbers $\mathbf{F}$ with entries
\begin{equation}
\bvec{F}_{\bvec{r}\bvec{q}} = 2 F_{\bvec{q}} \cos(\bvec{q} \cdot
\bvec{r})\qquad \bvec{r}\in {\mathcal{G}}\quad \bvec{q}\in \Lambda^{+}.
\label{eqn:fourierMatrix}
\end{equation}

Imposing positivity at our set of grid points, the minimum
charge principle immediately leads to a linear optimization problem:
\begin{eqnarray}
  & {\rm minimize:}\quad    & F_{\bvec{0}} \label{eqn:obj}\\
  & {\rm subject\ to:}\quad & F_{\bvec{0}}+\bvec{F}\cdot \bvec{s} \geq 0
  \label{eqn:cons}\\
  &                   & s_{\bvec{q}} \in \{-1,1\} \label{eqn:signs}
\end{eqnarray}
This type of problem is referred to in the optimization
literature as a mixed--integer program (MIP)
\cite{ref:wolsey,ref:MIPcomment}: the objective function and
constraints are linear in the unknowns, while some of the unknowns
are required to take discrete (integer) values. This general type of
problem has a long history in
operations research, with applications to scheduling, job assignment,
and set partitioning to name a few examples\cite{ref:NW}.
The traveling salesman problem is an example of
an integer program for which very large instances have recently been
solved. The minimum charge formulation of the centrosymmetric phase problem
most closely resembles a particular class of MIP known as the
multidimensional 0--1
knapsack problem, a $\mathcal{NP}$--hard combinatorial
optimization problem \cite{ref:mincharge,ref:knapsack}.

Even for a modest--sized problem like i(AlPdMn), the search tree for
500 measured reflections (unknown signs)
has $2^{500}$ leaves, so search by enumeration is impossible.
Several algorithms have been developed in an attempt to solve
problems of the kind specified by Eqns.~(\ref{eqn:obj})--(\ref{eqn:signs}).
One such algorithm, known as ``branch and bound'' (BnB), has been
used to solve some of the largest traveling salesman problems to
date \cite{ref:deutsch}. The basic idea of BnB is to use
information about bounds on the objective function to reduce the search
space.
\begin{figure}[t!]
\begin{center}
\includegraphics[width=3in]{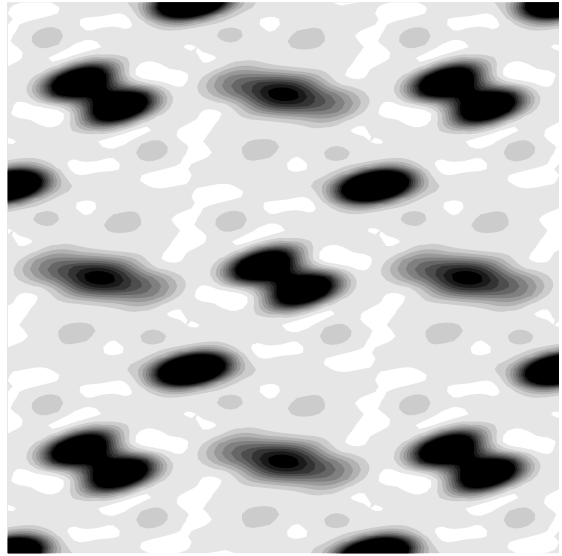}
\narrowtext{\caption{Reconstructed electron density in the 5-fold
plane of the 6D unit cell. Darker regions correspond to higher
charge density and can be identified with the three types of
atomic surface proposed by the authors of Ref.
17.\label{fig:atomicsurf}}}
\end{center}
\end{figure}
The feature of integer and mixed integer programming
problems that makes them so difficult is that the optimization is
over a nonconvex set. If we were to relax the integer constraints
on the signs and allow $s_{\bvec{q}} \in [-1,1]$, we have a related
problem called a linear program (LP). Linear programs are
\emph{not} $\mathcal{NP}$--hard so there exist a variety of
polynomial time algorithms to solve them \cite{ref:LP}.  Though
the LP for the phase problem is unphysical, LP solutions of
integer and mixed integer programs (known as ``relaxations'')
serve as upper (lower) bounds on the unrelaxed integer programs
whose goal is maximization (minimization).  We may use this
information to reduce the search space in the following way.
Suppose we are at depth $d$ in the tree, meaning we have fixed
the first $d-1$ signs at $\pm 1$ values.  Suppose also we have some
information about the upper bound on the charge, whether it be via
feasible though non-optimal solutions, or through \emph{a priori} knowledge.
We then solve two relaxed
LP problems; one in which $s_{d} = 1$ and one for which
$s_{d} = -1$, the relaxation occurring on all signs with index
greater than $d$.  If one of the relaxed problems yields a value
for the charge that is greater than our current best value, we
know the solution \emph{cannot} lie in that portion of the tree
and we may prune that entire branch.
\begin{figure}
\begin{center}
\includegraphics[width=3in]{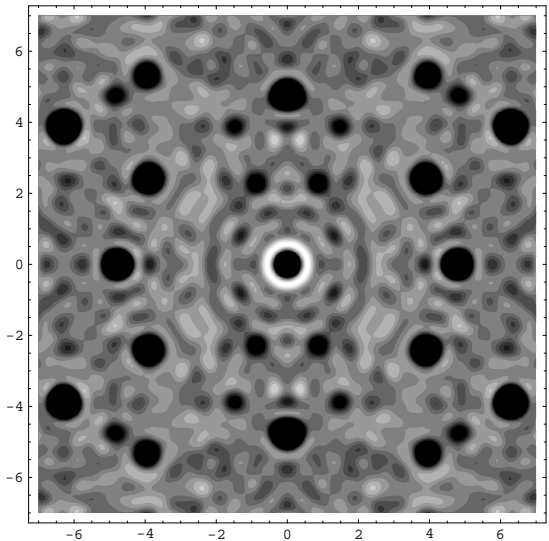}
\narrowtext{\caption{Contour plot of the i(AlPdMn) electron
density within an icosahedral mirror plane centered on the Mackay
cluster (a center of icosahedral symmetry).  The darker regions
correspond to regions of higher electron density.  Pairs of atoms
at unphysically short separation are manifestations of partial
site occupation in the average structure. The units on both axes
are \AA. \label{fig:fivecontour} }}
\end{center}
\end{figure}
The BnB search strategy that has proven most effective is a variation
of the ``breadth--first'' search, with the branching order specified by the
Fourier amplitudes. Thus signs corresponding to large amplitudes are
branched on first, while the very low amplitude signs remain
relaxed throughout the optimization. A basic parameter of the search
is an imposed upper bound on the charge. Beginning at the root, the
tree is searched exhaustively by branching on successively smaller
amplitude signs until the LP relaxations exceed the imposed upper bound.
This limits the depth of search into the tree, with deeper searches
corresponding to higher imposed bounds. Our experience with i(AlPdMn)
reproduces the behavior seen in studies with simulated diffraction
data, where a particular branch of the search tree is explored much
deeper than all other branches (given our imposed bound on the
charge). Were one to increase the imposed bound, eventually this
distinguished branch would be the first to reach the top of the tree
and yield the solution of the optimization problem. This is hardly
necessary, however, since the important characteristics of the
solution (large amplitude signs) have already been determined when the
distinguished branch first appears.

The efficiency of the algorithm is
improved by reducing the number of constraints in Eqn.
(\ref{eqn:cons}).
Since each inequality is associated with a point of the grid $\mathcal{G}$, we
wish to choose the smallest $\mathcal{G}$ consistent with the Fourier sampling
theorem which applies to our particular space group. If $G$ is the
corresponding point group, Eqn. (\ref{eqn:fourierMatrix}) generalizes to
\begin{equation}
\bvec{F}_{\bvec{r}\bvec{q}} = {F_{\bvec{q}}
\over{O(G_{\bvec{q}})}}\sum_{g\in G}
\cos(\bvec{q}_{g} \cdot \bvec{r}+\Phi(\bvec{q}_{g})) ,
\end{equation}
where $\Phi$ is the phase function of the space group\cite{ref:mermin}
and $O(G_{\bvec{q}})$ is
the order of the subgroup of $G$ that fixes any particular element of
the orbit, $\bvec{q}_{g}$. The columns of $\bvec{F}_{\bvec{r}\bvec{q}}$ thus
correspond to symmetry orbits of reflections. We make $\bvec{F}$
square and invertible by choosing the number of grid points to equal
the number of reflection orbits in the data set (plus one additional
point for $\bvec{q}=0$). The locations of the grid points within the
fundamental region of the space group are determined by making
$\bvec{F}$ well--conditioned. We use a simple procedure which starts
with random locations and performs conjugate gradient displacements
that maximize $\log\det \bvec{F}^{\rm t} \bvec{F}$\cite{ref:fourierGrid}.

We now present our results for the structure of i(AlPdMn), with space
group $F\bar{5}\bar{3}{{2}\over{m}}$.  Due
to space constraints we give a detailed report elsewhere
\cite{ref:detailedAlPdMn}. The minimum charge phase determination was
performed with 499 of the 503 symmetry inequivalent reflections in the
data set of Boudard et al.\cite{ref:boudard}. The deepest BnB search went to
depth 84 and required 2730 LP evaluations. A distinguished branch
already appeared in shallow searches with depth as small as 10. As a
6D structure i(AlPdMn) is relatively simple and the phase
determination effort is not so much that of \emph{finding} the solution but
more the process of refining (extending) the solution to the weaker
reflections.

To minimize on interpretation we only give the
reconstructed electron density, that is, we do
not attempt to identify atomic surfaces with idealized polyhedra and
specific chemical
occupation domains. Also, we note that since our electron density is derived
from Bragg reflection data, it represents the \emph{average} structure and
may therefore possess unphysically short interatomic distances, or
other manifestations of disorder. Strong indications of partial/mixed
occupational disorder were already found prior to phase determination in the
process of normalizing the experimental data\cite{ref:henley00}.

\begin{figure}
\begin{center}
\includegraphics[width=3in]{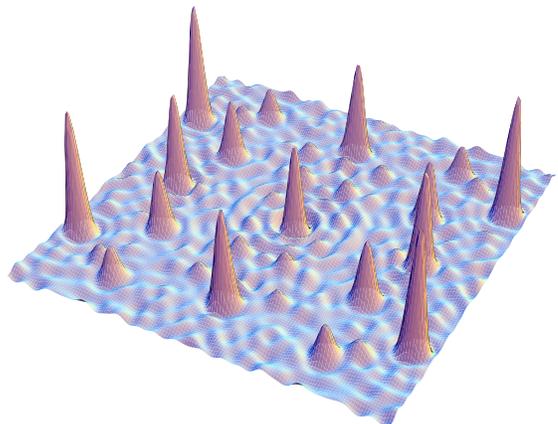}
\narrowtext{\caption{Height representation of
the electron density in Fig.~\ref{fig:fivecontour}.
\label{fig:five3D}
}}
\end{center}
\end{figure}

The main features in the electron density are seen in a ``5-fold
plane'', Figure \ref{fig:atomicsurf} --- a two dimensional section
that is commensurate with the 6D
lattice and invariant with respect to a 5-fold rotation. The compact
concentrations of charge in this plane are consistent with the three
atomic surface model derived by Boudard et al.\cite{ref:boudard} using
Patterson analysis. There is also some evidence in our reconstructed
``surfaces'' of features that go beyond the naive tiling models of
quasicrystalline structure. For example, deviations from a straight
``cigar'' shape indicate context dependent shifts from ideal tile
vertices.

By far the most conspicuous feature of the reconstructed atomic
surfaces is their rounded character. We believe this correctly
reflects the strong damping of intensities with the perpendicular
component of the 6D wave vector and is not an artifact of Fourier
truncation. A definitive test will require a larger x-ray data set.
The rounding of the atomic surfaces, particularly in the three
dimensions perpendicular to physical space, is expected in the random
tiling scenario\cite{ref:randomtile}, where the average structure is
effectively a
superposition of many tile rearrangements. Boundaries of the averaged atomic
surfaces (in the perpendicular dimensions) would not be sharp leading
to numerous partially occupied sites and apparent unphysically short
interatomic separations. Our reconstructed electron density in a
mirror plane through the origin in physical space,
Figs.~\ref{fig:fivecontour} and \ref{fig:five3D}, illustrates these
effects. Well defined atomic positions occur in the second shell and
correspond to the Mackay cluster. The sites of the first shell, at
vertices of a small dodecahedron, are only partially occupied to
avoid short interatomic separations.

We have described a new algorithm for direct method phase
determination based on optimization ideas from operations research.
Our application of the method to i(AlPdMn) shows the importance of
statistical disorder in the structure.  We are currently studying
the complexity of phase retrieval in general, including the
effects of diffraction data resolution.
In addition, we are applying the method to other instances of phase
retrieval,
including two--dimensional images and structure determination of
small biological molecules.

\acknowledgments We thank M.~P.~Teter for help in data
normalization.  K.~S.~Brown is supported by an NSF graduate
research fellowship.  A.~A.~Avanesov was supported under NSF
grant DMR-9820543 and DMR-MRSEC-9632275.

\end{multicols}


\begin{references}
\bibitem{ref:annealcomment}
SHELX \cite{ref:sheldrick} uses a stochastic annealing procedure
\cite{ref:anneal} while SnB uses effectively zero temperature
annealing, though the merits of one versus the other are unclear.
\end{references}
\end{document}